**Effect of cholesterol on the mechanical stability of gel-phase phospholipid bilayers studied by AFM force spectroscopy**

*Salomé Mielke[1] Raya Sorkin[2]* and Jacob Klein[1]*

1. Department of Molecular Chemistry and Materials Science, Weizmann Institute of Science, 76100 Rehovot, Israel

2. School of Chemistry, Raymond & Beverly Sackler Faculty of Exact Sciences, Tel Aviv University, 6997801, Tel Aviv, Israel

* Corresponding authors: email: rsorkin@tauex.tau.ac.il; jacob.klein@weizman.ac.il

We dedicate this article to Fyl Pincus, whose pioneering achievements in soft matter and biological physics, as well as his leadership and generosity of spirit over 5 decades, have given so much to the field

**Abstract**

The remarkably low sliding friction of articular cartilage in the major joints such as hips and knees, which is crucial for its homeostasis and joint health, has been attributed to lipid bilayers forming lubricious boundary layers at its surface. The robustness of such layers, and thus their lubrication efficiency at joint pressures, depends on the lipids forming them, including cholesterol which is a ubiquitous component, and which may act to strengthen of weaken the bilayer. In this work, a systematic study using an Atomic Force Microscope (AFM) was carried out to understand the effect of cholesterol on the nanomechanical stability of two saturated phospholipids, DSPC (1,2-distearoyl-sn-glycero-3-phosphatidlycholine) and DPPC (1,2-dipalmitoyl-sn-glycero- phosphatidylcholine), that differ in acyl chain lengths. Measurements were carried out both in water and in phosphate buffer solution (PBS). The nanomechanical stability of the lipid bilayers was quantitatively evaluated by measuring the breakthrough force needed to puncture the bilayer by the AFM tip. The molar fractions of cholesterol incorporated in the bilayers were 10% and 40%. We found that for both DSPC and DPPC, cholesterol significantly decreases the mechanical stability of the bilayers in solid ordered (SO) phase. In accordance with the literature, the strengthening effect of salt on the lipid bilayers was also observed. For DPPC with 10 mol % cholesterol, the effect of tip properties and the experimental procedure parameters on the breakthrough forces were also studied. Tip radius (2 - 42 nm), material (Si, $Si_3N_4$, Au) and loading rate (40 - 1000 nm/s) were varied systematically. The values of the breakthrough forces measured were not significantly affected by any of these parameters, showing that the weakening effect of cholesterol does not result from such changes in experimental conditions. As we have previously demonstrated that mechanical robustness improves the tribological performance of lipid layers, this study helps to shed light on the mechanism of physiological lubrication.

1. Introduction

Articulating joints exhibit extremely low friction, with friction coefficients of 0.002-0.02 up to pressures as high as 18 MPa (1). Several mechanisms have been proposed to explain this remarkable tribological performance, while at low speed and high load conditions the prevailing mechanism is that of boundary lubrication, as a result of fluid squeeze out from the gap of the joint (2). In this mechanism, the surface layers, rather than the properties of the lubricating fluid, are responsible for the friction reduction (3,4). One group of macromolecules prevailing on the cartilage surface is that of phospholipids, which have been associated with physiological lubrication for many years (5–7). Phosphatidylcholine (PC) is the predominant phospholipid component found on the cartilage surface, constituting 41% of the total phospholipid content. This composition includes various fatty acids with differing chain lengths and degrees of saturation. Within the PC content, both DPPC and DSPC make up approximately 30% (8,9). (8,9). We have previously studied the lubrication properties enabled by these lipids at physiologically high pressures (10,11) and demonstrated exceptionally low friction coefficients. However, while phospholipids by themselves are capable of friction reduction, at physiological conditions, the situation is more complex. Many other components are present, and a synergetic activity of these components might be responsible for the unique tribological properties of articulating cartilage (3,12). In this context, cholesterol is another major lipid component that is prevailing in the human body. Following our previous studies on friction reduction at physiologically high pressures by various phospholipid surface layers (11,13–16), and our discovery of the correlation between nanomechanical performance in the AFM and stability and lubrication in the SFB (10), we have set out to examine the nanomechanical properties of cholesterol-containing bilayers, which is the focus of this paper.

Cholesterol is found in eukaryotic membranes at concentrations around 20-30%, and even as high as 50% and 70% in certain cell types. It has many functions, being a precursor to hormones and vitamins, and also providing mechanical strength and controlling the phase behavior of membranes. Cholesterol incorporation in phospholipid monolayers and bilayers can change the ordering of the lipid chains (17). Within the gel phase, cholesterol can either increase or decrease the thickness of the bilayer depending on the chain length of the lipid. For saturated phosphatidylcholines containing 12-16 carbons per chain, cholesterol increases the thickness of the bilayer as it reduces the chain tilt. However, cholesterol reduces the thickness of 18 carbon chain bilayers below the phase transition temperature as the long phospholipid chains must deform or kink to accommodate the significantly shorter cholesterol molecule (18). It can thus be expected that cholesterol should increase the mechanical stability of 16 carbon chain SO bilayers and decrease the mechanical stability of 18 carbon chain SO bilayers. The effect of cholesterol on DSPC has not been examined so far, whereas AFM force spectroscopy studies of DPPC bilayers yielded contradicting results, demonstrating either decrease (19) or increase (20,21) in the mechanical stability of DPPC bilayers due to cholesterol addition. At cholesterol concentrations up to 25%, cholesterol rich (LO) and cholesterol poor (SO) phases coexist (22,23), and this can sometimes be observed as height difference of ~0.5nm in AFM scans(21).

In physiological conditions, lipid bilayers are not surrounded by pure water but by a salt solution. The most abundant, biologically relevant ions include $Na^+$, $K^+$, $Cl^-$, $Ca^{2+}$, and $Mg^{2+}$. Several studies suggest that cations induce a formation of lipid-ion networks, due to strong binding to particular sites on a lipid headgroup (24–28). Such sites may include the phosphodiester oxygens, or upon deeper penetration, the carbonyl group. Fukuma et. Al (26) suggest that the negatively charged phosphate groups share the positively charged cations, which results in an attractive electrostatic force exerted on the PC headgroups. The ion-lipid network formation results in a compression of the membrane coupled to an enhanced ordering of hydrocarbon lipid chains. This also leads to a notable decrease in the area per lipid. In addition to this, the cation−lipid complexation is responsible for a dramatic decrease in the lateral mobility of lipids(28–30). For example, for POPC bilayer with NaCl salt, a drop up to 50% in the diffusion coefficient was measured. The increase in the lateral interaction between the phospholipid molecules due to the lipid-ion network formation results in a more efficient packing, and thus is expected to increase the global mechanical stability of the membrane. Indeed, the forces required to penetrate PC bilayers with an AFM tip were found to increase with the increase in ionic strength (31,32).

In the presented work we aim to examine the effect of cholesterol addition to solid-ordered (SO) lipid bilayers on their mechanical stability. The effect of salt addition on these bilayers is tested as well, due to its physiological relevance. The breakthrough forces required to penetrate bilayers of varying cholesterol content are measured and compared. The measurements are performed by AFM force spectroscopy, a commonly used method for nanomechanical characterization of lipid bilayers (20,21,31,33–39). DSPC and DPPC bilayers with cholesterol molar percentage of 10% and 40% were examined. Contradicting nanoindentation results were reported earlier for cholesterol-containing DPPC bilayers (19–21). To shed light on the dispute, we examine in detail the experimental procedures and thereafter study the effect of the various experimental parameters on the results. We conduct measurements in water and salt solution, we vary the tip velocity systematically and we use three different cantilevers with different radii and materials for the force measurements.

## 2 Materials and experimental methods

### 2.1 Preparation of liposomes

The lipids 1,2-dipalmitoyl-sn-glycero-3-phosphatidylcholine (DPPC, 16:0) and 1,2-distearoyl-sn-glycero-3-phosphatidylcholine (DSPC, 18:0) were purchased from Lipoid (Ludwigshafen, Germany). Cholesterol was purchased from Sigma-Aldrich (St. Louis, MO, USA). For the preparation of small unilamellar vesicles (SUVs), standard approaches were used(13,40) [31, 32]. Briefly, phospholipids were mixed with the appropriate amount of cholesterol (to obtain either 10 or 40 molar percent) and dissolved in a 3:1 chloroform methanol solution. Next they were dried under nitrogen flow overnight, followed by drying in a desiccator. The dried lipids were dispersed in water and bath sonicated for 10-20 min at the appropriate temperature above the main phase transition of each lipid in order to obtain dispersed multilamellar vesicles (MLVs). Next, the multilamellar vesicles were progressively downsized by extrusion (Northern lipid Inc, Burnaby, BC, Canada) through polycarbonate filters with pore sizes of 400 nm, 100 nm and 50 nm. For each filter five extrusion cycles were performed. The liposomal size distribution (by volume) was determined in pure water using a viscotek 802 DLS. The water used throughout both for the liposome preparation and subsequent measurements was highly-purified (so-called conductivity) water from a Barnstead NanoPure system, with total organic content (TOC) ca. 1 ppb and resistivity 18.2 M$\Omega$.

**2.2 Preparation of lipid bilayer coated mica surfaces**

Bilayer covered mica surfaces were prepared as follows. Freshly cleaved mica (mounted in a petri dish (Falcon, 60 mm) was placed in a 0.3 mM SUV liposome dispersion prepared with Barnstead purified conductivity water. The mica surface was incubated overnight with the liposome dispersion at room temperature to ensure fusion of the liposomes and formation of bilayers. Thereafter the surfaces were rinsed to remove excess material by placing them in a beaker containing 200 ml water for 20 min. In order to examine the effect of salt on the mechanical properties of bilayers, liposome covered mica surfaces were prepared as described above. After rinsing with water, the water in the petri dish was replaced by a PBS solution (Sigma-Aldrich).

**2.3 Atomic Force Microscopy**

Imaging was done in tapping (AC) mode using a silicone nitride V-shaped 120 μm long cantilever having a nominal spring constant of 0.36±0.04 N/m with a pyramidal silicon tip with a radius of ~2 nm (SNL, Bruker). To get a good overview of the sample, images of different scan sizes (1-10 μm) and at different locations on the sample were taken. For the force plots, in most of the cases (unless otherwise specified) the same kind of tip as for the imaging (SNL, Bruker) was used. New tips were used for force mapping in order to avoid contamination of the tip as a result of the imaging procedure.

Prior to performing force measurements, the cantilever normal spring constant was obtained using the thermal calibration method (41). Force plots were captured in relative trigger mode using a trigger force of 5-30 nN. In order to obtain a large number of force plots at different locations on the sample, the function of force maps was used: In an area of 2x2 or 5x5 µm, 16x16 force plots were obtained each run, so each force map produced 256 force plots. For each system studied, several force maps were acquired at different locations. Unless otherwise specified the approach velocity of the tip was 400 nm/s. The data was analyzed with a MATLAB program that automatically detects the breakthrough forces of each force plot in a given force map. For all force maps about 5-10% of the data was analyzed manually and compared to the results of the program to validate the accuracy of the program.

## 3 Results

### 3.1 Characterization

Figure 1 shows AFM scans for the DSPC and DPPC bilayers studied in this work with 10% and 40% cholesterol content. All surface layers have a similar morphology of bilayer patches, as cross-sections through the layers (shown in the insets in Fig 1) indicate layers that are about 5-6 nm high. This is slightly higher than cholesterol-free bilayers, measured in previous studies to be 4-5 nm high(10,20). This is most likely due to thickening of the bilayer as a result of cholesterol addition, which is known to have an effect of bilayer tilt elimination (18). For DPPC with 40% cholesterol in PBS, the holes are only ~ 2.5 nm high (figure 1 (F)). There are two possible explanations: Either the holes are too small so that the tip is unable to reach the underlying surface, or these are one monolayer deep holes. No significant difference in the thickness of 10% and 40% cholesterol DSPC bilayers was observed. Also, no substantial difference in the morphology of the various systems was observed. We comment that generally, the size of the holes increased upon decrease in the liposome concentration in the incubation dispersion.

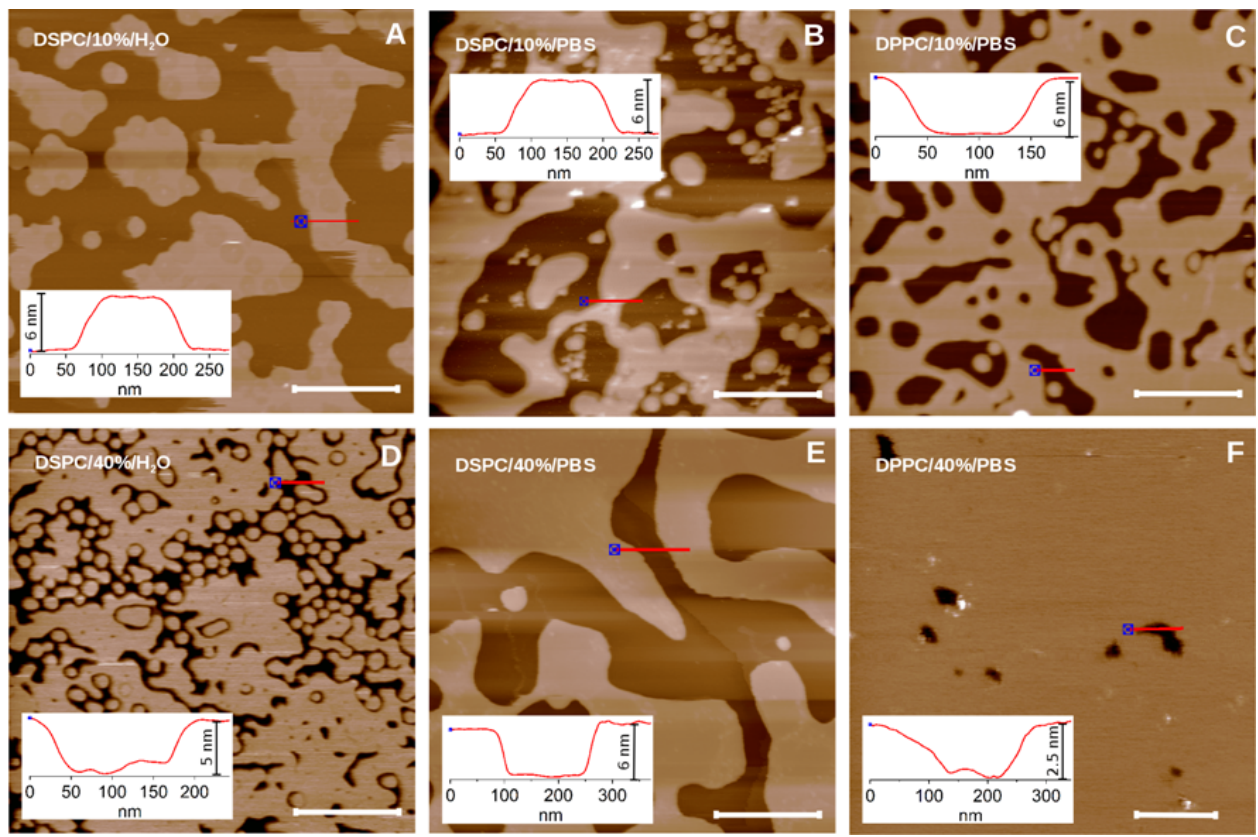

Figure 1. AFM images of bilayers. The thickness of the bilayers is shown in cross sections in the insets. All scale bars are 500 nm. (A) DSPC bilayer with 10% cholesterol in water. (B) DSPC bilayer with 10% cholesterol in PBS. (C) DPPC bilayer with 10% cholesterol in PBS. (D) DSPC bilayer with 40% cholesterol in water. (E) DSPC bilayer with 40% cholesterol in PBS. (F) DPPC bilayer with 40% cholesterol in PBS. Error in height measurements of uniform layers is 0.2 nm (the average standard deviation of the height measured in a uniform 0.5x0.5 um square).

**3.2 Force spectroscopy**

**3.2.1 DSPC bilayers in water**

Typical force plots for the different bilayers tested are shown in figure 2, with the corresponding histograms of the results shown in figure 3. Figure 2 (A) shows three representative force plots for a DSPC bilayer with 10% cholesterol. When summarized in a histogram (figure 3(A)), a bimodal distribution with $F_{b1} = 1.4 \pm 0.2$ nN and $F_{b2} = 4.3 \pm 0.8$ nN is observed. This is very similar to the values and distribution measured for pristine DSPC bilayers (10). The forces are significantly lower for a DSPC bilayer containing 40% cholesterol, as can be seen in figure 3 (D), where the forces are 0.2 ± 0.1 nN and 0.6 ± 0.1 nN. This is an indication that cholesterol addition (above 10%) decreases the mechanical strength of the DSPC bilayers. In addition, for DSPC with 40% cholesterol there is a substantial decrease in the scatter of the data, as seen in the histograms (when comparing figure 3 (A) and (D)). One possible explanation is that these bilayers are more homogeneous: For 10% cholesterol concentration, there should be segregation to cholesterol rich and cholesterol poor areas (21–23). For 40% cholesterol, the whole sample is cholesterol rich and therefore more uniform.

### 3.2.2 DSPC-chol bilayers in PBS

The histograms of the breakthrough forces for DSPC with 10% and 40% cholesterol in PBS are shown in figure 3 (B), (E). It can be seen that similarly to the water system, the breakthrough force for DSPC with 10% cholesterol is much higher than for DSPC with 40% cholesterol. Therefore, the general effect of cholesterol in reducing the mechanical stability of the bilayer is not affected by salt addition. When comparing the breakthrough forces of water and PBS systems, the forces measured in PBS are one order of magnitude higher than those measured in water. The enhanced stability of lipid bilayers due to addition of salt is likely explained by strong binding between cations and carbonyl oxygens, forming tight ion-lipid complexes with increased phospholipid-phospholipid lateral interaction (25,26,28,29,31), accompanied by salting out of hydration water, as bound ions will often occupy membrane binding sites for water and thus modify membrane affinity for water. Such ions replace or repel some of the bound water molecules and cause partial membrane condensation (42).

### 3.2.3 DPPC bilayers in PBS

DPPC was only examined in PBS and not in pure water, as the forces measured in salt are higher, and the differences in the magnitude of forces are more pronounced. Figure 2 (B) shows three representative force plots for a DPPC bilayer with 10% cholesterol, and a comparison between force plots obtained for DPPC in PBS, DSPC in PBS, and DSPC in water, all containing 10% cholesterol, is shown in figure 2(C). Clearly, DSPC bilayers are stronger then DPPC bilayers (higher breakthrough forces), and the presence of salt has a strengthening effect on the bilayer in both cases. Histograms of the breakthrough forces for DPPC with 10% and 40% cholesterol in PBS are shown in figure 3 (C), (F). Similarly to DSPC, DPPC also exhibits decrease in the breakthrough force with the increase in the cholesterol content.

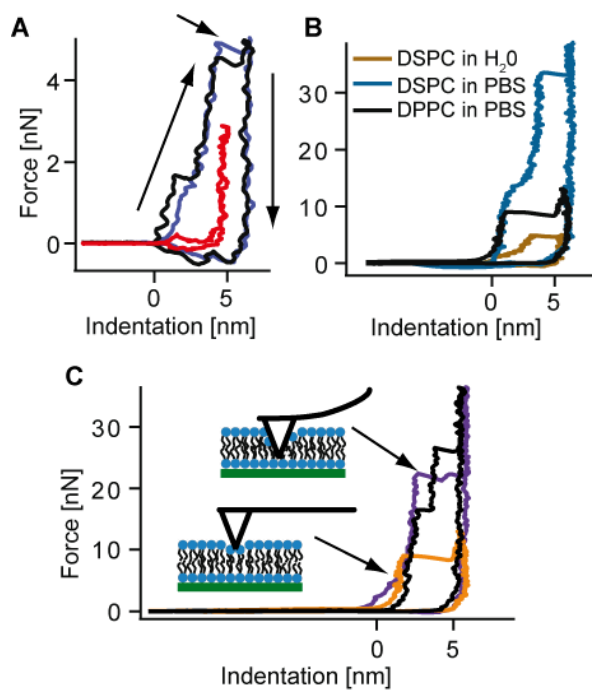

Figure 2. Force vs. indentation plots that show the penetration event. (A) Typical force vs. indentation plots for DSPC with 10% cholesterol in water. The three curves show the typical behavior: Most plots show only one penetration event with different breakthrough forces (low breakthrough force, red and high breakthrough force blue). Some plots show two penetration events (black). The arrows indicate the movement of the tip approaching the surface and moving backwards and the penetration event. (B) Typical force vs. indentation plots of DSPC with 10% cholesterol in water, DSPC with 10% cholesterol in PBS and DPPC with 10% cholesterol in PBS (as shown in the legend box). (C) Typical force vs. indentation plots of DPPC with 10% cholesterol in PBS. The three curves show the typical behavior: There are curves with two penetration events (black and purple) and curves with only one penetration event (orange). The two penetration events can have similar or different breakthrough forces. More detailed examination of the area under the of the area under the curves (to be addressed in future work) can also shed light on the work done on the bilayer by the tip**.**

The percentage of force plots exhibiting penetrations, out of all force maps recorded is shown in supplementary figure S1. When comparing DPPC and DSPC systems with the same cholesterol content, measured in PBS, the percentage of force plots exhibiting penetration events is higher for DPPC. This correlates with the lower breakthrough forces for DPPC, which have a shorter acyl chain length then DSPC. This results in weakening of the molecular interactions that result from the hydrophobic effect. We want to stress that for both DPPC and DSPC, we observed a weakening effect as a result from increase in cholesterol content. This is in contrast to the prediction that cholesterol should increase the strength of a DPPC bilayer and decrease the strength of a DSPC bilayer (43,44), as expected from the temperature shifts observed by DSC.

We have also examined the effect of tip approach velocity and tip radius and material on the mechanical strength of DPPC bilayers (see supplementary material) and found that these parameters did not change the trend of weakening bilayer mechanical strength upon cholesterol addition (see SI).

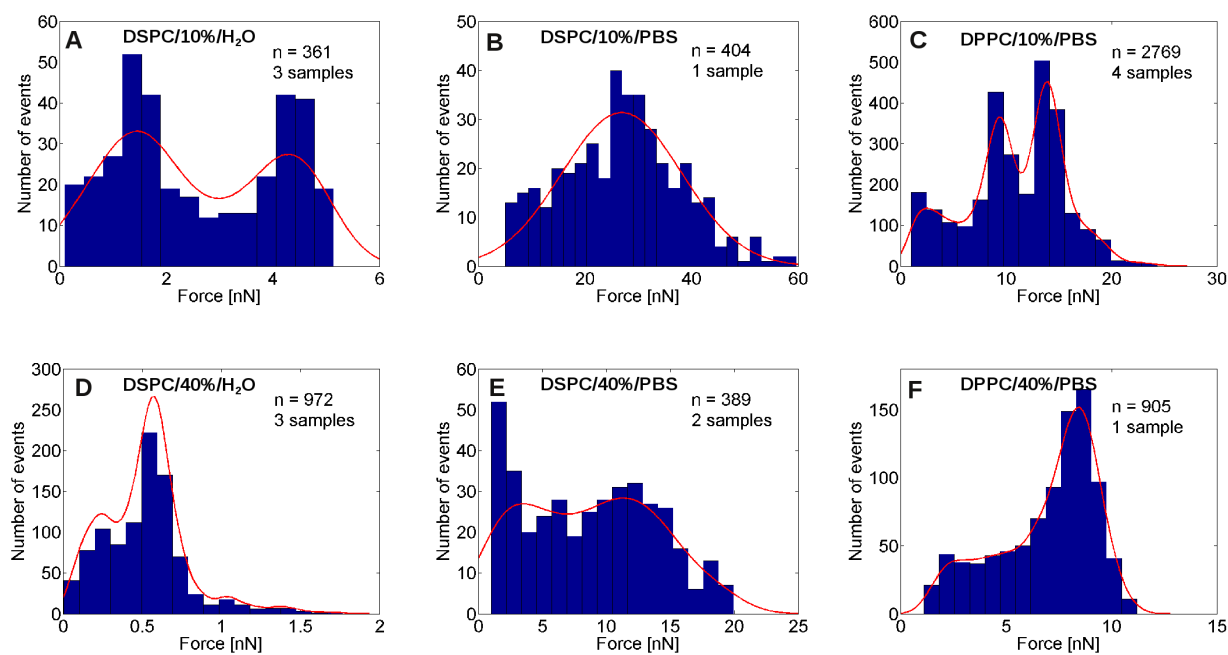

Figure 3. Histograms of breakthrough forces for different lipid bilayers. (A) DSPC bilayer with 10% cholesterol in water, a bimodal distribution is observed with breakthrough forces $1.4 \pm 0.2$ and $4.3 \pm 0.8$ nN (B) DSPC bilayer with 10% cholesterol in PBS, the breakthrough force is $26.9 \pm 11.2$ nN. (C)

DPPC bilayer with 10% cholesterol in PBS, a bimodal distribution is observed with breakthrough forces 9.3 ± 1.8 and 14.0 ± 2.0 nN. (D) DSPC bilayer with 40% cholesterol in water, a bimodal distribution is observed with breakthrough forces 0.2 ± 0.1 and 0.6 ± 0.1 nN. (E) DSPC bilayer with 40% cholesterol in PBS, the breakthrough force is 8.7 ± 5.1 nN. (F) DPPC bilayer with 40% cholesterol in PBS, the breakthrough force is 8.5 ± 1.0 nN. The inset shows the number of samples used to create the histogram and the total number of penetration events for each histogram.

## 4. Discussion

In this work we aimed to examine the effect of cholesterol addition to bilayers on their mechanical strength. The effect of cholesterol addition to DSPC bilayers on their mechanical properties has not been examined so far, whereas AFM force spectroscopy studies of DPPC bilayers yielded contradicting results, demonstrating either decrease (19) or increase (20,21) in the mechanical strength of DPPC bilayers due to cholesterol addition. McMullen et al. (1993) studied the transition temperature of bilayers with added cholesterol and showed that phospholipids that are shorter than cholesterol, which is equivalent to a 17-carbon acyl chain, display an increase in their transition temperature, whereas PC lipids with acyl chains longer then 17 carbons exhibit a decrease in their transition temperature. DPPC has 16 carbons per chain, whereas DSPC has 18, therefore the effect of cholesterol on their mechanical properties was expected to differ, in accordance with the difference in the effect cholesterol has on their transition temperature. This hypothesis was also mentioned by Spink et al. (1996). In our study, however, we observed a similar trend of decrease in mechanical stability upon cholesterol addition, for both DPPC and DSPC bilayers. We note that in the paper by McMullen et al. (1993), the fluctuations for DPPC and DSPC temperature shift data (fig. 8 in their paper) are of the same order as their difference, so that their data does not show a sharp transition in the trend for a lipid with 17 carbons per chain. In the future it would be meaningful to compare a shorter PC lipid, e.g. DLPC (12:0) that has 12 carbons per chain with a longer chain PC lipid that has 20 carbons per chain. In addition, DSC measurements of the transition temperatures of lipid mixtures at the compositions and conditions used in this study will shed further light at the behavior of these lipid mixtures. Nanoindentation measurements of pure DPPC in water, DSPC in water, DPPC in PBS, and DSPC in BPS should be performed in the future.

Our main finding is that the breakthrough force required to penetrate through a DSPC bilayer in the SO phase decreases with the increase in the incorporated cholesterol content. This trend was not affected by the presence of salt in the medium, and was revealed both in water and PBS immersed samples. We found the same trend for DPPC, also measured at room temperature in SO phase, in PBS. Interestingly, the effect of salt on the breakthrough force seems stronger than the effect of cholesterol. An increase in bilayer penetration force with the increase in ionic strength has been previously reported (24–28). It can be attributed to formation of an ion-lipid network, resulting from strong binding between ions and particular sites on a lipid headgroup (24–28), leading to enhanced lipid chain ordering. Concomitantly, this leads to a dramatic decrease in the lateral mobility of lipids(28–30). Further, such ions replace or repel some of the bound water molecules and cause partial membrane condensation (42). We attribute the reduced mechanical stability of both DPPC and DSPC gel bilayers to the perturbation by the cholesterol of the hydrophobic interactions between the acyl chains of the lipids and a consequent disruption to the ordered nature of the SO phase. This is in line with several studies utilizing electron spin resonance (ESR)(45,46) [47], [48], Raman(47,48) [49], [50], Fourier transform infrared spectroscopy (FT-IR) (49) [51] and $^2$H-nuclear magnetic resonance (NMR)(50) [52] methods, which led to a conception that cholesterol reduces molecular order in phospholipid membranes below the SO-LD phase transition temperature $T_m$, while enhancing molecular order at temperatures above $T_m$. The question of the effect of cholesterol on molecular ordering is however not universal and depends on the nature on the particular lipid, rather than simply on its degree of saturation (51,52). We note that the question of how cholesterol affects the bending rigidity of membranes which has been extensively studied for many years (52) and raised recent controversy (53,54) is intentionally not addressed here, as nanoindentation probes a different mechanical aspect of membranes.

Several histograms of the breakthrough forces presented in figure 3 show a bimodal behavior. A bimodal distribution for cholesterol containing bilayers has been observed before (21,55), and explained by coexistence of cholesterol rich and cholesterol poor areas in the bilayer that can sometimes be seen in the AFM scans (21,55). All our measurements were performed on samples without a visible phase separation. We note that pristine bilayers without added cholesterol also exhibited bimodal breakthrough force distribution, which was attributed to occasional intermittent adsorption of bilayers to the tip (10). Therefore, we believe that loss of bimodality in the breakthrough force values for DSPC 40% chol, compared with DSPC 10% chol, does not result from less significant phase separation. Rather, as bimodal $F_b$ values were recorded even for pristine DSPC bilayers, we attribute it to weaker tip-bilayer adhesion which suppresses the intermittent attachment of bilayers to the AFM tip (thereby eliminating the bimodal distribution).

When comparing force histograms of DPPC 10% chol in PBS measured with NP tip ($Si_3N_4$, r ≈ 20 nm) to a histogram obtained with an SNL tip (Si, r ≈ 2 nm) (fig S4 (A) and (B)), a clear bimodal distribution is observed with the sharp SNL tip. It seems, therefore, that resolution between two areas of different mechanical stability within the sample depends on the tip sharpness (all measurements in the work described here were done with a sharp SNL tip). Even though no clear segregation to two mechanically different phases was observed, the differences in the overall mechanical properties of the different systems is very clear; the breakthrough force for DSPC 10% chol is an order of magnitude higher then with 40% cholesterol. The weakening effect of the cholesterol is even more obvious in PBS, as the forces are higher. It is therefore very clear that an increase in cholesterol content results in a decrease in bilayer mechanical stability. These results are useful for understanding the tribological properties of cholesterol-containing bilayers and liposomes in lubricating boundary layers. They shed light on factors contributing to the robustness of such lipid-based biolubrication and thus for understanding of friction reduction mechanisms in synovial joints, as well as in the design of better treatment strategies of joint disorders, such as intra-articular injections of suitable lipid mixtures (3,56).

## 5. Summary and conclusions

In this work the influence of cholesterol on the mechanical properties of phosphatidylcholine bilayers was studied. The systems examined were DSPC and DPPC containing 10% and 40% cholesterol, immersed in pure water or PBS. The nano-mechanical stability of the bilayers was assessed by measuring the force needed for an AFM tip to penetrate through the membrane. Our main finding is that for both PC lipids, cholesterol significantly decreases the mechanical stability of the bilayers. Furthermore, we confirmed that the AFM tip radius and material did not have substantial influence on the measured breakthrough forces.

**Data awailability statement:**

Data will be made available on reasonable request

**Acknowledgements**:

We are grateful to Nir Kampf for useful discussions. We thank the European Research Council (advanced grant CartiLube 743016), the McCutchen Foundation, and the Israel Science Foundation (grant 1229/20) for financial support. This work was made possible partly through the historic generosity of the Perlman family.

**Author contributions:**

R.S., and J.K., designed research, S.M., performed research, S.M., R.S and J. K. wrote the paper.